\documentclass[%
aip,
rsi,%
amsmath,amssymb,
reprint,%
]{revtex4-1}

\usepackage{graphicx}
\usepackage{dcolumn}
\usepackage{bm}

\begin{document}
	
	\title{Photonic neuromorphic computing using vertical cavity semiconductor lasers}
	
	\author{Anas Skalli}
	\email{anas.skalli@femto-st.fr}
	\affiliation{FEMTO-ST Institute/Optics Department, CNRS \& University Bourgogne Franche-Comt\'e, \\15B avenue des Montboucons,
		Besan\c con Cedex, 25030, France.}

	\author{Joshua Robertson}
	\affiliation{Institute of Photonics, SUPA Department of Physics, University of Strathclyde, Glasgow, United Kingdom.}

	\author{Dafydd Owen-Newns}
	\affiliation{Institute of Photonics, SUPA Department of Physics, University of Strathclyde, Glasgow, United Kingdom.}

	\author{Matej Hejda}
	\affiliation{Institute of Photonics, SUPA Department of Physics, University of Strathclyde, Glasgow, United Kingdom.}

	\author{Xavier Porte}%
	\affiliation{FEMTO-ST Institute/Optics Department, CNRS \& University Bourgogne Franche-Comt\'e, \\15B avenue des Montboucons,
	Besan\c con Cedex, 25030, France.}

	\author{Stephan Reitzenstein}%
	\affiliation{echnical University Berlin/Institute of Solid-State Physics and the Center of Nanophotonics, Hardenbergstra{\ss}e 36, D-10632 Berlin, Germany.}

	\author{Antonio Hurtado}
	\affiliation{Institute of Photonics, SUPA Department of Physics, University of Strathclyde, Glasgow, United Kingdom.}

	\author{D. Brunner}
	\affiliation{FEMTO-ST Institute/Optics Department, CNRS \& University Bourgogne Franche-Comt\'e, \\15B avenue des Montboucons,
		Besan\c con Cedex, 25030, France
	}%

	\date{\today}
	
	\begin{abstract}
		
Photonic realizations of neural network computing hardware are a promising approach to enable future scalability of neuromorphic computing.
The number of special purpose neuromorphic hardware and neuromorphic photonics has accelerated on such a scale that one can now speak of a Cambrian explosion.
Work along these lines includes (i) high performance hardware for artificial neurons, (ii) the efficient and scalable implementation of a neural network's connections, and (iii) strategies to adjust network connections during the learning phase.
In this review we provide an overview on vertical-cavity surface-emitting lasers (VCSELs) and how these high-performance electro-optical components either implement or are combined with additional photonic hardware to demonstrate points (i-iii).
In the neurmorphic photonics' context, VCSELs are of exceptional interest as they are compatible with CMOS fabrication, readily achieve 30\% wall-plug efficiency and >30~GHz modulation bandwidth and hence are highly energy efficient and ultra-fast.
Crucially, they react highly nonlinear to optical injection as well as to electrical modulation, making them highly suitable as all-optical as well as electro-optical photonic neurons.
Their optical cavities are wavelength-limited, and standard semiconductor growth and lithography enables non-classical cavity configurations and geometries.
This  enables excitable VCSELs (i.e. spiking VCSELs) to finely control their temporal and spatial coherence, to unlock Terahertz bandwidths through spin-flip effects, and even to leverage cavity quantum electrodynamics to further boost their efficiency.
Finally, as VCSEL arrays they are compatible with standard 2D photonic integration, but their emission vertical to the substrate makes them ideally suited for scalable integrated networks leveraging 3D photonic waveguides.
Here, we discuss the implementation of spatially as well as temporally multiplexed VCSEL neural networks and reservoirs, computation on the basis of excitable VCSELs as photonic spiking neurons, as well as concepts and advances in the fabrication of VCSELs and microlasers.
Finally, we provide an outlook and a road-map identifying future possibilities and some crucial milestones for the field.		
	\end{abstract}
	
	\maketitle

\section{Introduction \label{sec:Intro}}

\begin{figure}[h]
	\begin{center}
		\includegraphics[width=0.75\linewidth]{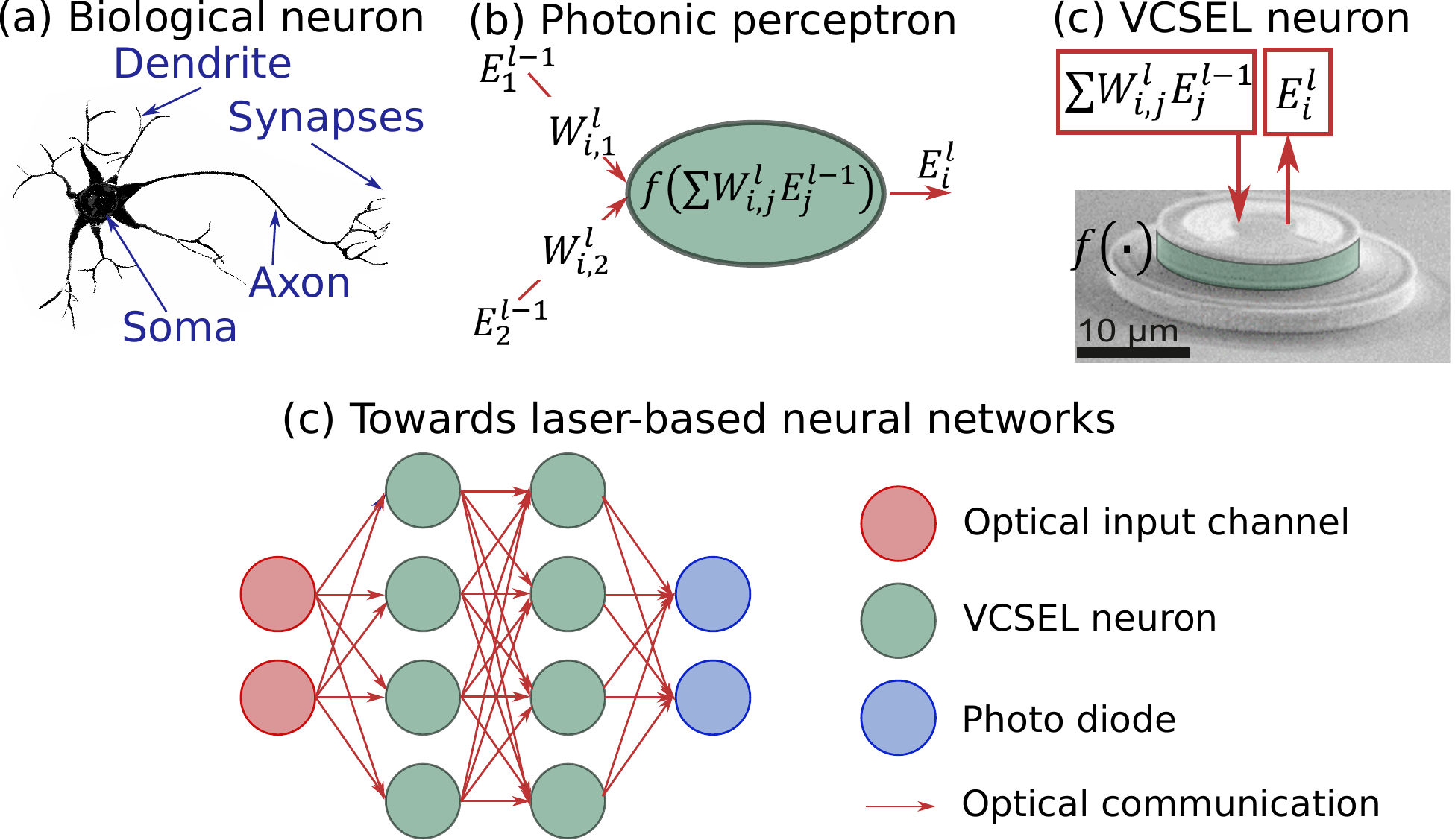}  
		\caption{(a) Biological neurons form complex connections using synapses, which link axons and dendrites of pre- and post-synaptic neurons.
			Crucial for computation is their nonlinearity.
			(b) The optical perceptron mimics this functionality by nonlinearly transforming a weighted sum of optical input fields.
			(c) VCSELs can efficiently be coupled to external fields, and their internal workings make them respond nonlinearly to such optical injection.
			(d) Linked via photonic connections, VCSELs offer a promising roadmap to photonc neural networks.}
		\label{fig:PNNconcept}
	\end{center}
\end{figure}

Neural networks (NN) are concepts fundamentally relying on a \emph{connectionist} approach to computation.
This concept was developed based on the most-simplified features of a biological neuron, see Fig. \ref{fig:PNNconcept}(a).
In a biological neuron, an axon typically forms a long-range link that connects via synapses to the dendrites of the post-synaptic neuron.
Furthermore, a neuron's inner working principles makes them respond nonlinearly to inputs.
Following this principle, a NN creates and combines a large number of simple nonlinear transformations by artificial neurons, or perceptrons, while tweaking the network's topology during a training phase such that specific computations result from emergence.
This enables computing outside the classical symbolic programming and Boolean logic-gates catalogue: programming is replaced by statistical NN-topology optimization, while the hardware-concept is based on networks of simple nonlinear units rather than on mostly serial threads of logic gates.
The increased flexibility in programming a computer enables addressing highly abstract computational challenges, and NNs currently drive a revolution in various areas of economy, technology and society.
However, in a NN concept, all artificial neurons express their transformations of input signals simultaneously, which intimately links NNs to parallelism.

Today's computing substrates can follow the connectionist principle only within tight limits, and the development of scalable, efficient, high performance hardware NNs currently is a major goal.
The principle operations are information transformation, information transduction across the network connections, and topology optimization.
Leveraging photons is highly promising for implementing the NN's parallel interconnect \cite{Lohmann1996} and in particular in combination with photonic NN (PNN) computing \cite{Farhat1985,Shastri2021}.
In a photonic perceptron, see Fig. \ref{fig:PNNconcept}(b), optical inputs from the network are accumulated and nonlinearly transformed.
Optical communication is inherently parallel \cite{Psaltis1990}, energy efficient \cite{Miller2017} and recent developments in optical memristors indicate a road-map towards efficient programmability \cite{Feldmann2021}.

However, competitive nonlinear photonic components for artificial neurons have been a challenge.
Lately, novel devices move the associated energy required consumption ever closer to electronics \cite{Liu2015}, and a general strategy to enhance nonlinearity is to confine photons into a tight space or to increase the interaction time, often by using optical resonators such as the case for lasers.
Vertical-cavity surface-emitting lasers (VCSELs) are one of the most prominent semiconductor lasers, with unique properties making them highly suitable for next generation PNN hardware, see Fig. \ref{fig:PNNconcept}(c) presenting an excerpt of a VCSEL during fabrication. 
VCSELs profit from high-yield and commercially mature fabrication, readily reaching above 30\% wall-plug efficiency \cite{Haghighi2021}.
Furthermore, due to their low lasing threshold currents and amplitude-phase coupling via Henry's alpha factor, they react highly nonlinear to optical input and can therefore act as all-optical as well as electro-optical artificial neurons.
VCSELs can readily be modulated with >30 GHz \cite{Haghighi2021}.
Such high efficiency and high speed combined result in an ultra low energy per nonlinear transformation on the order of 10 fJ \cite{Heuser2020a}.
Considering the parallelism of a potentially passive and low-loss massive interconnect comprising $>10^2$ connections per channel, this brings PNNs using VCSELs into the realm of <100 aJ per Operation, compared to 100~fJ$\dots$1~pJ in electronic circuits.
Crucially, this is only a current snapshot, and spin-VCSELs \cite{Lindemann2019} as well as high-$\beta$ VCSELs leveraging cavity quantum electrodynamics can further reduce this cost towards the fundamental physical limits on the order of a few photons per operation \cite{Wang2021}.

VCSELs emit vertically to their substrates, which allows for efficient testing and for interfacing with scalable 3D integrated photonic circuits \cite{Moughames2020} or external optical resonators \cite{Brunner2015} in order to establish a PNN's connections, see Fig. \ref{fig:PNNconcept}(d).
Furthermore, in order to realize numerous photonic neurons they can be arranged in arrays \cite{Heuser2020a}, or, as recently demonstrated, one can leverage spatial multiplexing of a multimode large area VCSEL in order to implement photonic neurons in spatial modes \cite{porte2021complete}.
An additional concept to extend the number of degrees of freedom are the two orthogonal polarization directions \cite{vatin2019experimental}.
Finally, VCSEL structures can be operated in an excitable regime, either relying on intra-cavity saturable absorbers \cite{Barbay2011} or on injection-locked induced excitability \cite{Hurtado2015}.

Owing to their excellent properties, semiconductor laser-based PNNs have a long standing history \cite{Mos1997}.
Initially, the difficulties in training and implementing large neural networks limited the advance of the field.
With the introduction of Reservoir Computing (RC) \cite{Jaeger2004} this challenge was efficiently mitigated, and in combination with a temporal-multiplexing approach \cite{Appeltant2011} large-scale PNNs could be successfully emulated \cite{VanderSande2017}.

In this review, we provide a detailed overview on the implementation of VCSELs in different PNN topologies and different photonic neuron concepts.
In Sec. \ref{sec:Time_Delay} we discuss the temporal-multiplexing approach for PNNs, in Sec. \ref{sec:Spatio_temporal} the implementation of PNNs in spatially extended multimode lasers, while excitable photonic spiking VCSEL neurons are discussed in Sec. \ref{sec:Excitable}, the fabrication of modern VCSELs structures in Sec. \ref{sec:Fabrication}.
At the end of our review we provide an extensive outlook in Sec. \ref{sec:Outlook}, which includes a roadmap to guide the field's future development.

\section{Principles of a VCSEL \label{sec:VCSEL_basics}}

Photonic hardware provides neuromorphic computing with low-loss optical interconnects, intrinsic non-linearity and ultrafast operation rates~\cite{Miller2017}.
Suitable devices include optical modulators and different semiconductor lasers such as edge-emitting lasers, resonant tunnelling diode-laser-diode systems and VCSELs~\cite{Prucnal2016,Tait2019}.
The latter are of particular interest because of their commercial availability, their compactness (in comparison to edge-emitters) and their vertical emission with close-to-ideal Gaussian emission profiles. 

\begin{figure}
	\begin{center}
		\includegraphics[width=0.5\linewidth]{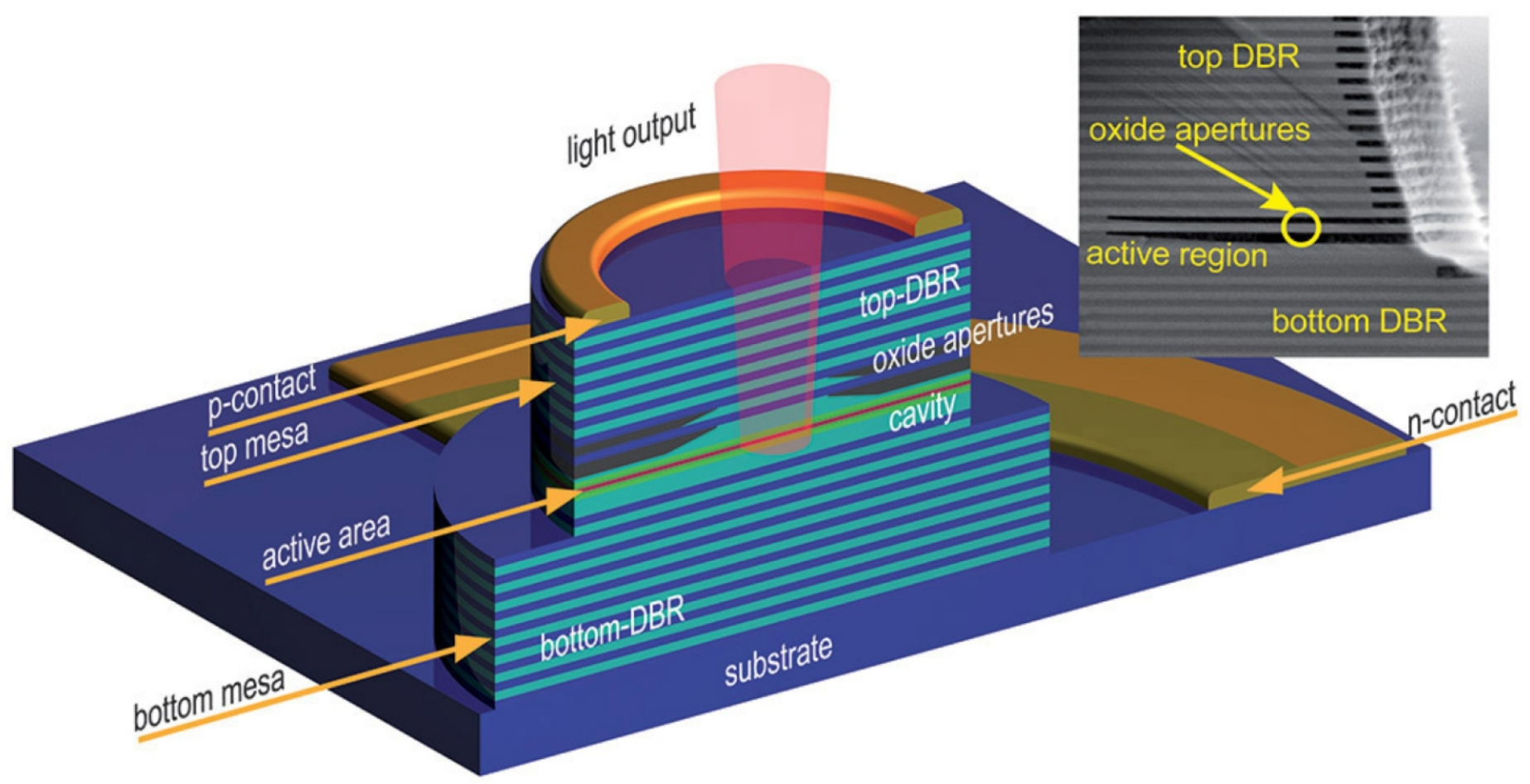}  
		\caption{Schematic representation of a VCSEL device. The compact, vertical emitting laser consist essentially of two doped DBRs, a central cavity with the active area, at least one oxide aperture, and ring-shaped n- and p-contacts. Inset: Scanning electron microscope image showing the central area (cross-section) of a VCSEL with two oxide apertures. Reproduced with permission from Ref.\cite{Liu2015}}
		\label{fig:VCSEL_scheme}
	\end{center}
\end{figure}

VCSELs can have many different configurations, yet they all feature identical design principles.
Horizontally arranged high-reflectivity dielectric mirrors sandwich a gain material inside the cavity formed between the two mirror.
Often, this cavity is very short and only half a wavelength in length.
These planar structures are then etched to form pillars, or mesas, and optical emission is through the top-mirror that typically features a slightly lower reflectivity ($\sim$99.8\%).
A schematic illustration of this principle is given in Fig. \ref{fig:VCSEL_scheme}, and details of semiconductor VCSEL growth can be found in Sec. \ref{sec:Fabrication}.
Mirrors are usually distributed Bragg reflectors (DBR), comprising layers of different semiconductor materials transparent at the lasing wavelength yet with different refractive indices, while optical gain amplification can either be realized using one or more quantum wells or ensembles of quantum dots.

The dynamics of VCSELs are usually described using rate equation models.
Compared to standard edge emitting semiconductor lasers, VCSEL's do however require a more complex description, as they can commonly emit in and hence require modeling of orthogonal optical polarizations and the associated carrier spin-populations.
The standard set of equations is based on the spin-flip model \cite{SanMiguel1995SFM}.
These features make their numerical description more complex, but this complexity is an assert for computation and allows for higher dimensionality or complexity of VCSEL-neurons and excitable behaviour, see Sect. \ref{sec:Excitable}. 

The most relevant parameters in such a description are the carrier's relaxation oscillations, whose usual timescale of 0.1$\dots10~$GHz typically determines the bandwidth of a VCSEL's nonlinear transformation.
Furthermore, if information is to be injected optically, then the usually $1\dots5~$ps lifetime of a photon within the VCSEL's cavity is important.
Optical information injection requires locking the VCSEL serving as photonic perceptron to an external injection laser \cite{Bueno2016Locking}.
Such optical injection locking can only be realized inside a narrow range of frequency detuning between injection and response laser.
The width of this detuning window is determined by the ratio of injection and the VCSEL's emission power, and the VCSEL's photon lifetime acts as a scaling factor: the longer the lifetime, the more frequency selective the VCSEL, and consequently the narrower the injection locking window is for a given power ratio.

\section{Time Delayed Reservoirs \label{sec:Time_Delay}}

\begin{figure}[t]
	\begin{center}
		\includegraphics[width=0.8\linewidth]{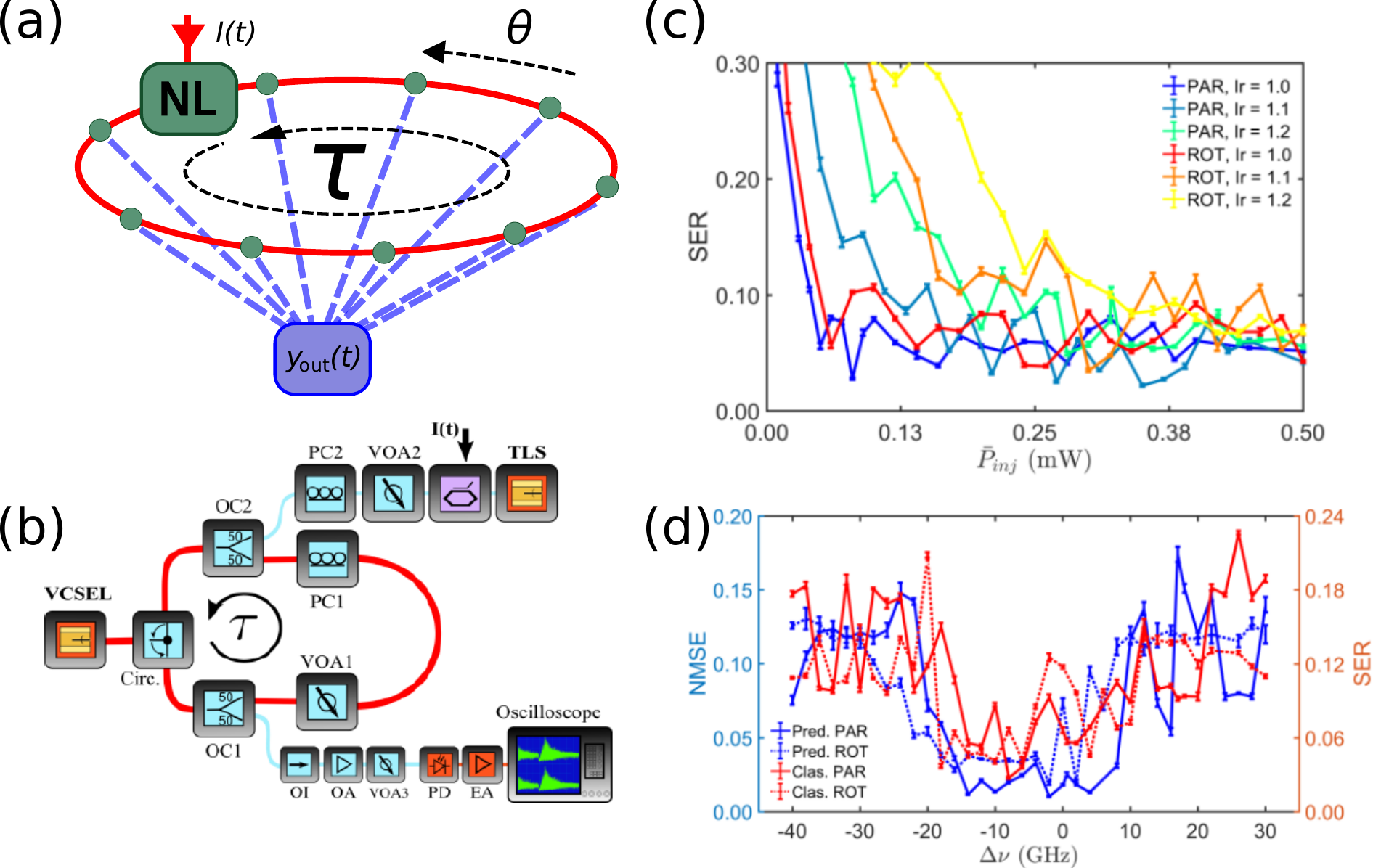}  
		\caption{(a) Delay-based Reservoir Computing. 
			(b) Scheme of the experimental setup for a VCSEL delay-based reservoir computer. OC: Optical Coupler, PC: Polarization Controller, VOA: Variable Optical Attenuator, TLS: Tunable Laser Source, PD: Photodiode, EA: Electrical Amplifier, OI: Optical Isolator (Fig. from~\cite{bueno2021comprehensive}).
			(c) Performance of a VCSEL delay RC on a classification task as a function of the average optical injected power for parallel (PAR) and rotated (ROT) feedback (Fig. from~\cite{bueno2021comprehensive}). 
			(d) Performance of the same VCSEL RC on prediction (Pred.) and classification (Clas.) tasks for PAR and ROT feedback as a function of the frequency detuning (Fig. from~\cite{bueno2021comprehensive}).}
		\label{fig:Delay_VCSEL}
	\end{center}
\end{figure}

The concept of delay-based reservoir computing was first introduced in a seminal paper of Appeltant et al.~\cite{Appeltant2011}. 
Mathematically, delay systems are described by delay differential equations whose temporal solution depends on the present as well as on past states. 
As such, when delayed feedback is added even a low-dimensional system offers the high-dimensional phase space which is the basis for RC. 
In contrast to spatially-extended RC systems, this approach uses only a single nonlinear node and a linear delayed feedback line with roundtrip duration $\tau$ as illustrated in Fig. \ref{fig:Delay_VCSEL}(a). 
The nonlinear node continuously transforms the information fed back from the delay line. 
A number of virtual nodes (or virtual neurons) are then created by discretizing the continuous output of the delayed system in $N$ time segments of length $\theta$, fulfilling the relation $N = \tau / \theta$. 
Delay RC processes the input information $I(t)$ sequentially and their operation speed is determined by the long delay length $\tau$. 
Each input sample is mapped into the complete set of virtual nodes $N$ and multiplied by a random mask $m(t)$ with periodicity $\tau$. 
This temporal mask is introduced to diversify the response of the virtual nodes to the input signal and plays a similar role as the input weights in a conventional reservoir. 
Inertia \cite{Appeltant2011} or de-synchronization between injection mask length and delay $\tau$ \cite{Paquot2012} realize the reservoir's internal connections.
Typically, the input masking and the output layer weights are implemented off-line with a digital computer. 
The values $y_{\mathrm{out}}(t)$ (cf. Fig. \ref{fig:Delay_VCSEL}(a)) are obtained one at a time in intervals of $\tau$ from the continuous analogue output, and they are calculated as a linear combination of the $\theta$-spaced states of the virtual nodes, creating reservoir state x(t). 
The samples are then concatenated to state matrix $S$ where columns represent time and rows represent the neurons. 
The content of $S$ is therefore determined by the input, the masking, and the dynamical behavior of the nonlinear node. 

Delay-based RC aims at minimizing the expected hardware complexity of spatially-extended systems using a single nonlinear node only, a delay line and high-bandwidth capabilities for injection and detection. 
When implemented in photonics, delay-based RC systems have successfully been employed for classification, prediction and system modelling tasks with state-of-the-art results. 
Specific arguments photonic delayed RC implementations is the availability of off-the-shelf fiber-optics telecommunications hardware and that the lengths of the delay line does not impact the modulation bandwidth $\theta^{-1}$, cf. Fig.~\ref{fig:Delay_VCSEL}(b). 
The first optical hardware implementations of RC were independently developed by Larger et al.~\cite{Larger2012} and Paquot et al.~\cite{Paquot2012} using optoelectronic systems. 
Soon after, the first implementation with a semiconductor laser showed excellent performance for RC at Gbyte/s rates~\cite{Brunner2013}, the system's injection locking parameter space was investigated in~\cite{Bueno2016Locking}. 

Implementations involving VCSELs leverage their polarization properties to improve the information processing performance and speed. 
Typically, one of the two optical modes dominates, but their relative intensities will change depending on laser, optical feedback and injection parameters. 
Time delayed RC based on VCSELs explore polarization dependencies to build alternative feedback schemes by rotating the polarization of the feedback or even exploring dual performance by simultaneously computing with the two modes. 
Vatin et al. numerically~\cite{vatin:hal-01877236} and experimentally~\cite{vatin2019experimental} first demonstrated delay-based RC with two-mode polarization dynamics of a VCSEL, which offers a bigger playground for optical feedback and injection configurations.
Using either parallel (PAR) or orthogonal (ROT) configurations, a comprehensive experimental analysis can be found in~\cite{bueno2021comprehensive}, where the authors analyze the performance of both feedback configurations with Mackey-Glass prediction and nonlinear channel equalization tasks, see Fig.~\ref{fig:Delay_VCSEL}(c,d). 
The authors find that ROT feedback degrades the computational performance in terms of memory capacity when there is a significant power difference between the two emission modes~\cite{bueno2021comprehensive}. 

Dual emission also offers the possibility of implementing two tasks in parallel, where each task is performed by one of the two perpendicularly polarized modes. 
This was first proposed numerically in~\cite{guo2019polarization} and later on demonstrated experimentally in~\cite{vatin2020experimental}. 
Dual task operation slightly degrades the system's performance for each task, but nevertheless represents an attractive advantage of VCSELs.
Finally, Harkhoe et al.~\cite{harkhoe2021neuro} proposed and numerically investigated the use of fast spin-flip dynamics in VCSELs to boost the information processing speed at multi-GSa/s. 
The speed of the spin-flip dynamics, which depend on the birrefringence of the VCSEL, can further be increased by more than one order of magnitude~\cite{Lindemann2019}.

\section{Spatio-temporal reservoirs \label{sec:Spatio_temporal}}

Delay-based reservoirs enable minimal hardware requirements \cite{VanderSande2017}, yet information is still processed sequentially, requiring the implementations of virtual, time multiplexed nodes \cite{Appeltant2011} and computational speed still depends on the size of a time-multiplexed PNN layer.
It also entails that most implementations still rely on an external computer to pre-process data, construct the reservoir state and calculate the output weights for training.
In \cite{duport2012all}, an online learning strategy is implemented to reduce the influence of the external computer.
However, the sequential nature remains a fundamental feature (and asset) of this system.

In contrast, by using large area VCSELs (LA-VCSELs), using a VCSEL with an aperture of $25~\mu\text{m}$ and leveraging their highly multimode nature, a truly parallel, spatially multiplexed, VCSEL-based reservoir was introduced in \cite{porte2021complete}.
Due to parallelism, computing speed does in this system not depend on the PNN's size any longer.
In addition, an online learning strategy was implemented, making the system's computation fully autonomous and relegating the external computer to a simple supervision and instrument-control role.

\subsection{Working principle and experimental setup}

\begin{figure}[h]
	\begin{center}
		\includegraphics[width=1\linewidth]{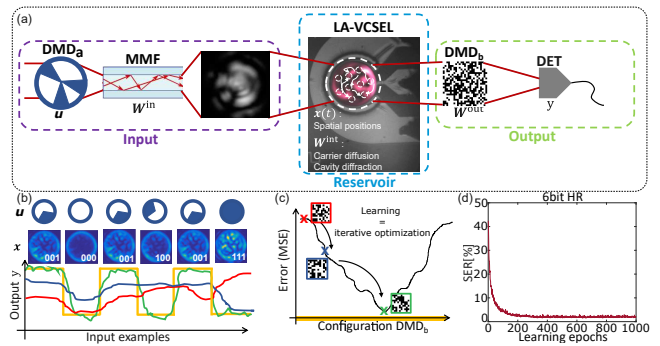} 
		\caption{(a) Working Principle of the LA-VCSEL spatially multiplexed reservoir.
			(b) Input information $\mathbf{u}$ and the LA-VCSEL response for 3-bit binary headers.
			The graph shows the target output $y^{\text{target}}$ and different reservoir outputs $y^{\text{out}}$ of decreasing mean square error (MSE) (red, blue and green).
			(c) Schematic illustration of the error landscape, showing the MSE as a function of the output weights configuration.
			The outlined (red, blue and green) Boolean matrices correspond to the output weights giving the output from b).
			(d) Representative performance of the PNN on a 6bit header recognition (HR) task. Reproduced with permission from \cite{SkalliPhysProp2021}.}
		\label{fig1_spatial}
	\end{center}
\end{figure}

The PNN implemented in \cite{porte2021complete} can be divided into three sections, c.f. Fig.~\ref{fig1_spatial}(a).
First, the input layer is realised via a digital micromirror device (DMDa) and a multimode fibre (mmf).
Spatial patterns (Boolean images) displayed on DMDa constitute the input information, $\mathbf{u}$ in Fig.~\ref{fig1_spatial}(a), and the mmf passively implements the PNN's input weights via its complex transmission matrix.

The second part is the reservoir itself.
The output field of the mmf is optically injected via imaging onto the LA-VCSEL top-facet, which implements all the components of the reservoir through its internal principles and nonlinear dynamics.
Nodes are spatially multiplexed positions on the surface of the LA-VCSEL, and coupling is taken care of by carrier diffusion and optical diffraction inside the LA-VCSEL’s cavity.
The reservoir state is then the perturbed mode-profile of the LA-VCSEL under optical injection, denoted in Fig.~\ref{fig1_spatial}(a) as $\mathbf{x}$.
In Fig.~\ref{fig1_spatial}(b), the VCSEL's response is shown for several 3-bit headers.
Responses to each input pattern are complex, different, and non-trivial.
This explains, in an intuitive sense, how finding a configuration of output weights that solve a certain computational task like header recognition, XOR, and digital-analogue conversion \cite{porte2021complete} is possible.
Ultimately, the reservoir is used at speeds orders of magnitudes below its inherent timescales, and the LA-VCSEL was operated in it's steady state, hence no recurrent properties such as fading memory of the device were exploited.

The last part of this PNN is its output layer.
To implement output weights $\mathbf{W}^{\text{out}}$, the LA-VCSEL's near field is imaged onto a second DMD (DMDb).
The reflection off DMDb in one direction is imaged onto a large area detector, and the mirrors of DMDb sample the different positions, i.e. neurons on the LA-VCSEL's surface.
This applies a Boolean weight matrix to the reservoir state, and the authors implemented $\sim$90 trainable Boolean readout weights. 
The output of the network $y^{\text{out}}$ was recorded at the detector for a set of input patterns called the training batch.
For each image in the training batch, the desired target output $y^{\text{target}}$ is known.
Thus, after each training epoch (a run of one batch), $y^{\text{out}}$ is recorded and a normalized mean square error is calculated $\epsilon_{k} =||y^{\text{out}}_{k}(t) - y^{\text{target}}||$.
Training is realised via a simple, yet effective evolutionary algorithm presented in \cite{bueno2018reinforcement,andreoli2020boolean}.
Boolean weights (mirrors) at random positions are flipped at the transition from $k$ to $k+1$.
If the change is beneficial, i.e. $\epsilon_{k+1} < \epsilon_{k}$, it is kept, otherwise the output weights are reset to the configuration at epoch $k$ as shown in Fig.~\ref{fig1_spatial}(b,c).
This operation is repeated until the desired performance threshold is met.
Figure~\ref{fig1_spatial}(d) shows a representative learning curve for a 6bit header recognition task, the system reaches around a $1.5\%$ symbol error rate (SER).

\subsection{LA-VCSEL PNN metrics}

\begin{figure}[t]
	\begin{center}
		\includegraphics[width=0.9\linewidth]{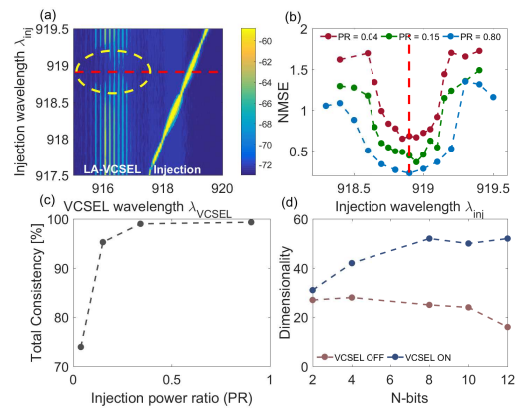}  \caption{a) Injection locking of the VCSEL by an external drive laser.
			b) Performance (NMSE) vs Injection wavelength for different injection power ratios (PR), highlighting that the best performance is reached for the injection locking conditions.
			c) Total system consistency as a function of $\text{PR}=P_{\text{inj}}/P_{\text{VCSEL}}$.
			d) Dimensionality of the system with the VCSEL ON and OFF for different bit-numbers.
			The VCSEL expands the dimensionality of the input highlighting the non-linearity of the device. Reproduced with permission from XXX.}
		\label{fig2_spatial}
	\end{center}
\end{figure}

The performance of the LA-VCSEL reservoir is extensively studied in \cite{SkalliPhysProp2021}, mapping the impact of several parameters on the overall performance.
Figure~\ref{fig2_spatial}(a) shows how the LA-VCSEL's free running modes react to external optical injection.
At $\lambda = 918.9~$nm, a resonance condition is met where the LA-VCSEL's free running modes are suppressed (by $\approx10~d\text{B}$) and the device locks to the injection laser, that is to say, the VCSEL's emission wavelength is shifted to that of the external drive laser.
Injection locking has been extensively studied in \cite{ackemann1999spatial,ackemann2000patterns}.
In Fig.~\ref{fig2_spatial}{}b), we see a clear dependence of the performance on the injection wavelength as well as the injection power ratio ($\text{PR} = P_{\text{inj}}/P_{\text{VCSEL}}$), and the best performance is obtained under locking, consisten with \cite{Bueno2016Locking} for a semiconductor laser delay reservoir.

In addition, the authors link consistency, i.e. the ability of the system to respond in the same way when subjected to the same input, and dimensionality (numbers of degrees of freedom of the system) to the physical parameters $\lambda_{\text{inj}}$, $P_{\text{inj}}$, and $I_{\text{bias}}$.
Figure~\ref{fig2_spatial}(c) shows the clear dependence of consistency on optical injection power, the total system consistency saturates and reaches an excellent level above $99.5\%$.

In the same paper, a general way to gauge a system's dimensionality was proposed.
The same random input sequence is sent onto the LA-VCSEL, and the response of every neuron is recorded.
The method relies on principal component analysis for noisy systems \cite{malinowski1977theory,turner2006noise}, and can generally be applied to other hardware ANNs.
In Fig.~\ref{fig2_spatial}(d), the dimensionality is measured for sequences of headers with the LA-VCSEL switched off as well as with the LA-VCSEL switched on.
It is clear that the LA-VCSEL significantly increases the dimensionality of data representation.
However, one has to say that rather than viewing the obtained dimensionality as absolute values, one should use their relative variation in direct comparisons, such as here to demonstrate the clear benefit of the LA-VCSEL on the PNN's capability to expand the dimensionality of data representation.
This is due to approximations made in \cite{malinowski1977theory,turner2006noise} and due to a slightly heuristic criteria for assigning principle components to noise.



\section{Computing with Spiking VCSEL Neurons \label{sec:Excitable}}

A certain isomorphism between VCSELs and biological neurons has started to be explored. Here, we introduce VCSEL-based systems that are able to directly mimic (at ultrafast rates) the spiking action potentials of biological neurons and use the resulting photonic spiking signals to process information.

Biological neurons are known to communicate using electrical action potentials which take the form of temporal spiking signals. These are created when a neuron is subject to a stimulation from an external source or by neighbouring neurons. Remarkably, the generation of optical excitable spiking signals has also been observed in VCSELs, at ultrafast sub-nanosecond rates, multiple orders of magnitude faster than the time-scales of biological neurons \cite{Hurtado2015,Barbay2011}. However, while the different nonlinear dynamical responses occurring in VCSELs, including excitability, have been widely reported in the literature, it is only in recent years that the link between the nonlinear effects in VCSELs and neuronal behaviour has been proposed \cite{Hurtado:10,Hurtado2012,AlSeyab2014,Robertson2017controlled,Deng2017,Deng2018,Robertson2020Toward,Turconi2013Control,Robertson2019,Garbin2018,Dolcemascolo2018} towards novel paradigms in neuromorphic spike-based photonic processing systems.

\subsection{Reports of excitability and neuronal responses in VCSELs}

A first experimental report outlining the use of VCSELs as artificial photonic neurons appeared as early as 2010 \cite{Hurtado:10}. That work described the use of optically-induced polarisation switching in a 1550\,nm VCSEL to trigger different types of nonlinear activation functions, reproducing the response of neurons to excitatory and inhibitory stimuli. In a subsequent report in 2012 \cite{Hurtado2012}, the link between the spiking dynamical responses triggered in VCSELs and those used to process information by neurons, was proposed. In that work, optical injection into the subsidiary (orthogonally-polarised) mode of a 1550\,nm VCSEL induced polarisation switching. The authors investigated the polarization resolved dynamics of the stimulated VCSEL and showed responses reproducing different neuronal dynamics, such as phasic spiking and tonic spiking. This was the first implementation of a dynamically excited VCSEL-neuron yielding neural-like responses at sub-ns time-scales, a method that was further theorised in \cite{AlSeyab2014} to explore the possibility of triggering excitable spiking responses in a VCSEL operating at the key telecom wavelength of 1550\,nm. It was found that around the injection locking boundary (for both parallel and orthogonally-polarised optical injection), different excitable responses could be activated. Like in biological neurons excitable spikes had an activation threshold requirement, therefore revealing an experimental method of implementing controllable artificial VCSEL-based spiking neurons. The system was subsequently realised experimentally \cite{Hurtado2015}, demonstrating that excitable neuron-like spiking dynamics could indeed be triggered in a 1550\,nm-VCSEL with precise control. As theorized, short 0.5\,ns input stimulations successfully triggered 100\,ps-long spikes using both orthogonal and parallel polarized optical injection, revealing also a high reproducibility of the spiking outputs. The precise control of neuron-like excitability was also demonstrated through the manipulation of the injection modulation. It was shown that increasing the duration of the stimulation triggered the continuous firing of tonic spikes. In subsequent experimental reports, the authors reported also the achievement of spike inhibitory behaviour \cite{Robertson2017controlled}, communication of excitable spikes between coupled VCSEL-neurons \cite{Deng2017,Deng2018} and their potentials for photonic spike memory operation and for the emulation of retinal neuronal circuits \cite{Robertson2020Toward}.

\begin{figure}[h]
	\begin{center}
		\includegraphics[width=0.95\linewidth]{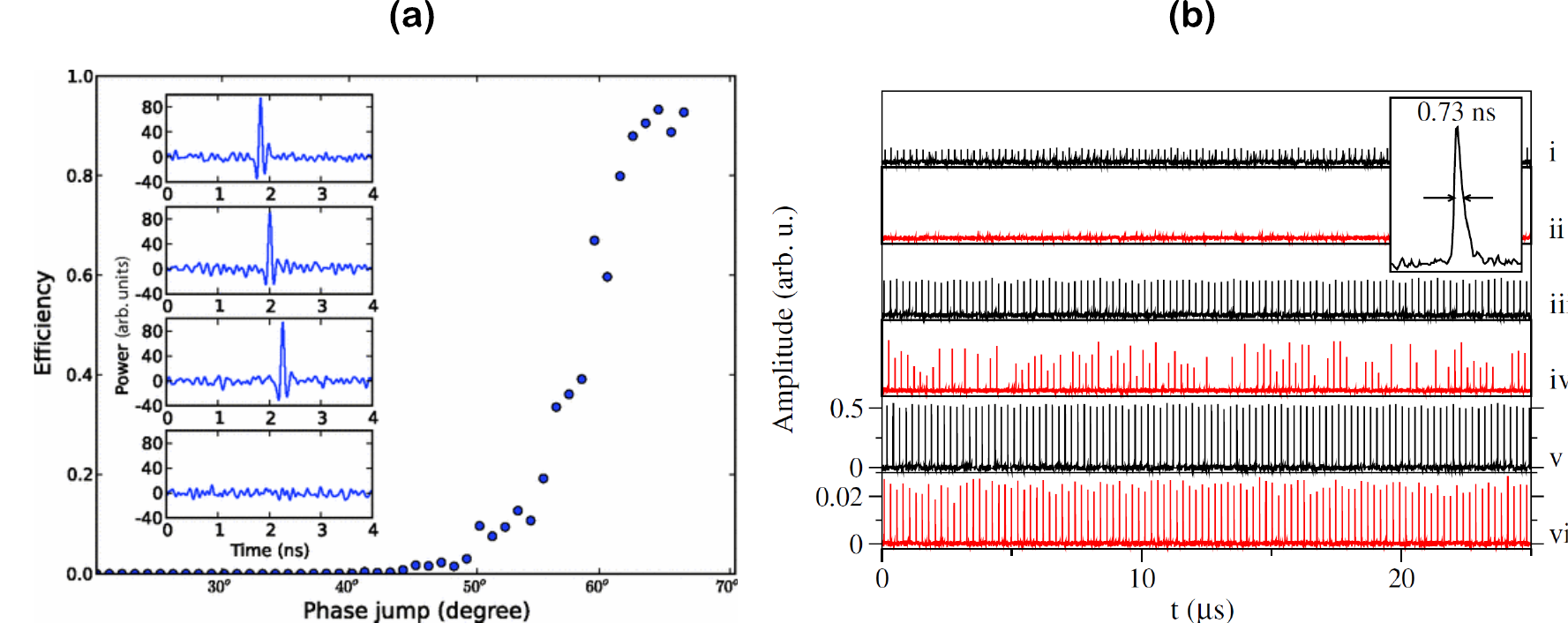} 
		\caption{Excitable spiking responses activated by a) phase jump modulations in an injection locked VCSEL \cite{Turconi2013Control} and b) optical pump modulations in a near threshold VCSEL-SA \cite{Barbay2011}. a) The efficiency curve for excitable responses is plotted with respect to injection phase jump amplitude. The insets show individual excitable responses for phase jumps of 66$^{\circ}$ (top), 55$^{\circ}$, and 52$^{\circ}$ for successful (third) and unsuccessful (bottom) cases. b) Time series show the consistency of 100 excitable responses (red) to input pulses (black) of 0.4 (i,ii), 1.0 (iii,iv) and 1.78 (v,vi) input strength (scaled to the excitable threshold). The inset shows a 730\,ps excitable response from the VCSEL-SA in series vi. Reproduced with permissions from \cite{Barbay2011,Turconi2013Control}}
		\label{fig:Excitable VCSELs}
	\end{center}
\end{figure}

Excitable pulses have also been observed in short-wavelength VCSELs under phase-modulated optical injection. M. Turconi et al. \cite{Turconi2013Control} reported excitable pulses in a 980\,nm VCSEL around the injection locking boundary when adding 100\,ps-long phase jumps of different amplitudes in the optically injected signals. Fig.\,\ref{fig:Excitable VCSELs}\,a) shows the shape and efficiency of the phase-triggered excitable responses. Again, the excitable pulses required inputs to exceed an activation threshold for successful and consistent firing. The authors also showed that strong pulses in bias current could also trigger excitable responses from the VCSEL. The electrically-triggered spiking responses, however, were found to be less reliable and longer (typically 1\,ns in duration) than those produced by optical injection with phase modulation. The activation of excitable spikes using the modulation of VCSEL bias current was again later demonstrated experimentally using VCSELs subject to intensity modulated optical injection \cite{Robertson2019}. Using phase-modulation, Garbin et al. \cite{Garbin2018} also reported the regeneration of excitable optical spikes in a short-wavelength VCSEL under delayed optical feedback towards photonic spike memory operation. More recently, experimental and theoretical results successfully demonstrated resonator and integrator behaviour in phase modulated VCSEL systems producing multipulse excitability \cite{Dolcemascolo2018}.  

In parallel, excitable spiking signals have also been investigated in VCSELs containing intracavity saturable absorbers (SA). We refer to these devices from now onwards as VCSEL-SAs. Work led by Barbay et al. \cite{Barbay2011} has focused experimentally on spiking micro-pillar VCSEL-SA systems, demonstrating that optically-pumped micropillar VCSEL-SAs (with emission at 980\,nm when optically pumped at 800\,nm) exhibited self-pulsating (spiking) regimes immediately after their lasing threshold. Barbay et al. also demonstrated the activation of 730\,ps-long excitable spikes under optical pump modulation (Fig.\,\ref{fig:Excitable VCSELs}\,b), and the existence of a spike activation threshold. In addition, this team also reported neuronal processing features in spiking micropillar VCSEL-SA structures, including spike firing latency, refractoriness and integration of multiple inputs prior to spike firing \cite{Selmi2014,Selmi2015,Selmi2016}. These spiking behaviours were further confirmed by numerical simulations using the Yamada model for a semiconductor laser with a SA region. Numerical work by these authors also outlined the potentials of networks of evanescently coupled micropillar VCSEL-SAs to perform temporal spike logic operations (e.g. OR, AND) \cite{pammi2019photonic}. 

The activation of neuronal dynamics in VCSELs have also seen significant theoretical investigation via the SFM and Yamada models \cite{Yamada1993,Miguel1995}. In addition to numerical work carried out by the teams at Strathclyde \cite{Hurtado2015,Robertson2017controlled}, Nice \cite{Dolcemascolo2018} and Paris \cite{Barbay2011,Selmi2016,pammi2019photonic}, other groups also started to investigate numerically the spiking properties of VCSEL-neurons. In 2016, S. Xiang et al. \cite{Xiang2016} reported numerical results based on the SFM, validating the early 2012 experimental results of \cite{Hurtado2012}, and later expanded on that experimental work to widely report theoretically on the potentials of dynamical polarisation switching responses in VCSELs for high-speed neuronal-like functionalities \cite{Xiang2017}. In addition, a great body of theoretical work has recently appeared in literature focusing on optical spiking neurons based upon VCSEL-SAs \cite{Nahmias2013,Skontranis2021,Zhang2018,Xiang2018,Zhang2018Polarization,zhang2020controllable,Xiang2018Numerical,Xiang2019stdp,song2020spike,Zhang2020Winner,Xiang2021Computing,xiang2020all}. An early theoretical work \cite{Nahmias2013} used a two-section rate-equation model to describe a VCSEL-SA and demonstrate numerically that excitable pulses could be activated in these devices under the injection of short optical pulses. This work also revealed that VCSEL-SAs could theoretically operate as a leaky integrate-and-fire (LIF) neurons, and operate in different interconnected architectures with brain-inspired connectivity. Multiple other theoretical works further investigated the spiking responses in VCSEL-SAs and their potential for a wide diversity of tasks and procedures, ranging from spiking convolutional neural networks for image processing \cite{Skontranis2021}, to spiking information encoding and storage \cite{Zhang2018,Xiang2018,Zhang2018Polarization,zhang2020controllable}, Sudoku solvers \cite{Gao2021}, unsupervised learning procedures based on Spike-Timing Dependent Plasticity (STDP) \cite{Xiang2018Numerical,Xiang2019stdp,song2020spike}, spike pattern recognition \cite{Zhang2020Winner,Xiang2021Computing}, spiking XOR gate implementations \cite{xiang2020all}, amongst others.

\subsection{Spike-based processing systems with VCSELs}

\begin{figure}[h]
	\begin{center}
		\includegraphics[width=0.95\linewidth]{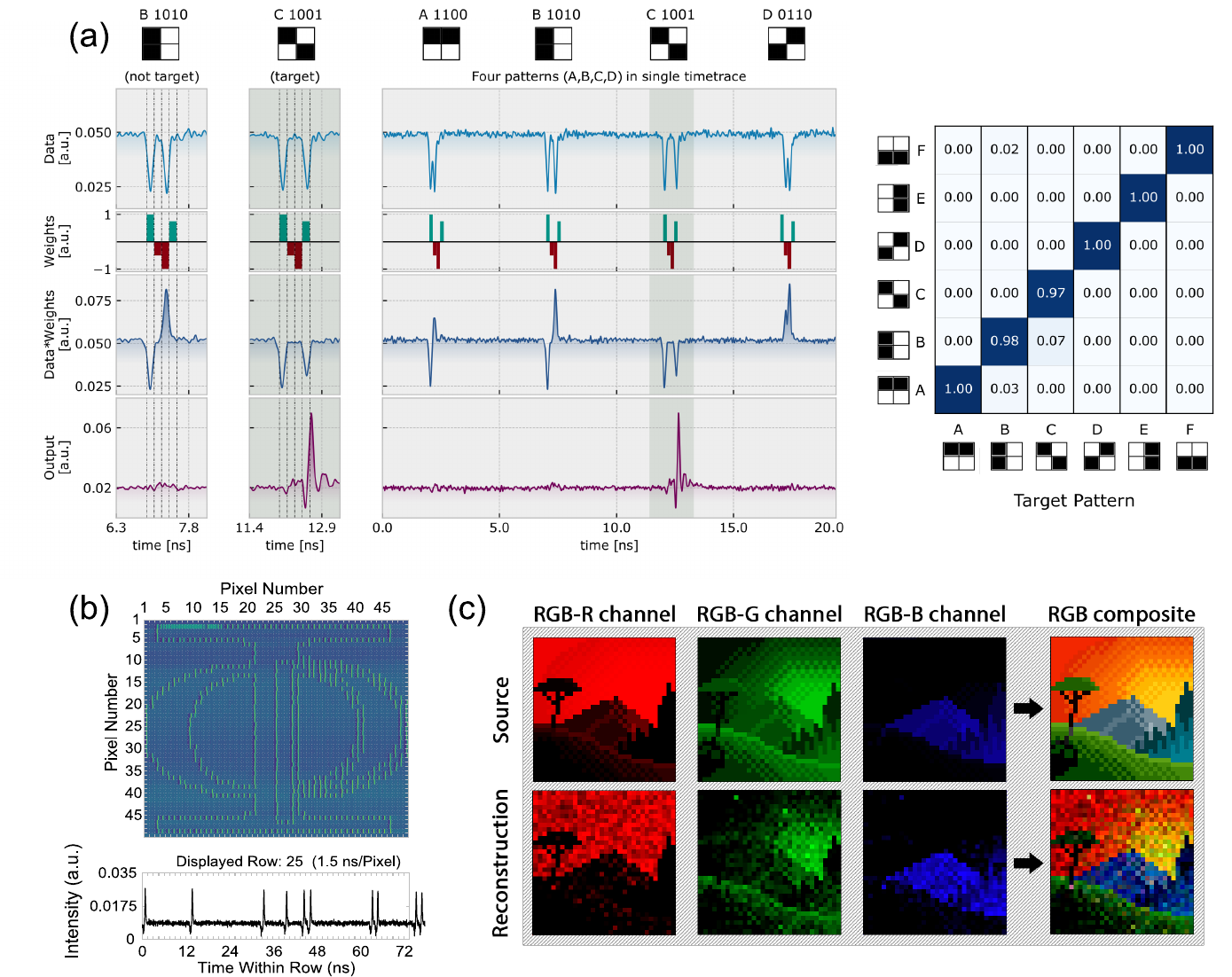} 
		\caption{Experimentally realised applications of excitable spiking VCSELs. a) Fast (100\,ps) input patterns are time division mulitplexed, weighted and injected into a VCSEL-neuron. Only the target input pattern results in the activation of an excitable response (a successful recognition). The confusion matrix reveals the recognition efficiency of various target patterns \cite{Robertson2021Ultrafast}. b) Gradient-based edge detection of a black and white image (50x50 pixel Institute of Photonics logo) via convolution with a combination of vertical and horizontal kernels. Excitable VCSEL responses (bottom) corresponds to row 25 of the image, revealing the activation 10 excitable spikes \cite{robertson2020image}. c) The fast (nanosecond rate) encoding and decoding of a RGB colour image into a spiking signal with variable spiking frequency. The RGB source image is compared the with spiking reconstruction after spike frequency analysis \cite{Hejda2021}. Reproduced with permission from \cite{robertson2020image,Hejda2021,Robertson2021Ultrafast}.}
		\label{fig:Excitable Applications}
	\end{center}
\end{figure}

It is only very recently that the first experimental demonstrators of spike processing systems, with VCSEL neurons for ultrafast neuromorphic photonic computing, have started to emerge \cite{robertson2020,Zhang2021Experimental,robertson2020image,zhang2021all,Robertson2021Ultrafast,Hejda2020,Hejda2020}. Using VCSEL spiking neurons, several processing tasks, typical of artificial intelligence (AI) implementations, have been reported (e.g. image processing, and temporal pattern recognition) with the desired advantages of ultrafast processing times and low energy usage. In 2020, Robertson et al. \cite{robertson2020} demonstrated experimentally a method of classifying 4-bit digital patterns using a single VCSEL-neuron. In that work, as shown in Fig.\,\ref{fig:Excitable Applications}\,a, 4-bit patterns (with a bit rate of $\sim$150\,ps) were injected optically into a VCSEL-neuron which fired fast 100\,ps spikes only in response to target patterns, remaining quiescent otherwise. In the same work \cite{robertson2020}, the authors used the leaky integrate-and-fire nature of a VCSEL-neuron to demonstrate experimentally a coincidence detection task, permitting the system to recognise the arrival of two different inputs within a very short temporal window ($<$420\,ps). Similarly, all-optical XOR classification has been demonstrated on binary patterns, emulating pyramidal neurons, using an optical injection system subject to dual modulation \cite{Zhang2021Experimental}.    

More recently, spiking VCSEL-neurons have also been applied for image processing. Experimental demonstrations of image edge-feature detection \cite{robertson2020image} and binary convolution (that calculates the intensity gradients in images) \cite{zhang2021all} with spiking VCSEL neurons have been reported.  Notably, these different functionalities were achieved at high processing rates (with GHz’s input data signals) and low input power levels (tens of $\mu$Ws), using hardware friendly systems and commercially available VCSELs. The latter reacted by triggering spikes (100\,ps long) for image-pixels where specific features were detected, as shown in Fig.\,\ref{fig:Excitable Applications}\,b for the logo of the University of Strathclyde's Institute of Photonics. Recent work by Robertson et al \cite{Robertson2021Ultrafast} has also shown that a single VCSEL-neuron can implement a full neuronal layer, performing image edge detection, by subsequently combining this photonic VCSEL layer with a software-implemented Spiking Neuronal Network (SNN). This hybrid VCSEL and software-implemented SNN demonstrated the accurate classification ($>$97\% precision) of complex images from the MNIST hand-written digit image database. These works showed spike processing rate of 1.5\,ns/pixel using a single commercially-sourced VCSEL without any specific device design optimisation. Higher processing speeds and more complex functionalities are expected for bespoke device designs and systems using multiple VCSEL neurons simultaneously.

Additionally, recent work by Hejda et al \cite{Hejda2020} demonstrated experimentally the capability of spiking VCSEL-neurons to act as spike information encoders, using different spike coding mechanism, such as precise spike timing and spike rate encoding. Hejda et al demonstrated experimentally that VCSEL-neurons, like their biological counterparts, show a spike latency and inter-spike refractory period that is dependent on input stimulation amplitude. This was used to implement precise spike-timing encoding of digital signals (digital to spiking format conversion) at rates over 1 Gbps, and spike rate encoding, where the amplitude of an strong (weak) input stimulus yielded faster (slower) spike firing frequencies. In a subsequent work, Hejda et al \cite{Hejda2021} capitalised on the biologically-inspired rated coding ability of VCSEL-neurons, to demonstrate fast (nanosecond rate) encoding of image colour information for image processing functionalities (Fig.\,\ref{fig:Excitable Applications}\,c).

Artificial VCSEL-based optical neurons therefore provide exciting platforms towards fast (GHz rates) and low-energy neuromorphic photonic spike-processing systems. With access to threshold-and-fire and integrate-and-fire functionality, spiking VCSEL-neurons could be integrated into large interconnected photonic SNNs  architectures for the implementation of complex light-enabled spike-based processing functionalities (e.g. image processing, computer vision, pattern recognition, etc.). VCSELs thus provide a consistent, all-optical, low power, and hardware friendly solution towards future ultrafast neuromorphic computing systems for AI technologies.

\section{VCSEL and microlaser fabrication\label{sec:Fabrication}}

The novel computing concepts discussed in this review rely crucially on vertically emitting lasers. VCSELs emit normal to the chip surface, can be controlled electrically, show high-speed dynamics and are energy efficient~\cite{Michalzik2012} and are much more compact than edge emitting lasers. In this regard, micropillar lasers are of high interest~\cite{Gies2019}. These nanophotonic structures with diameters in the few-$\mu$m range allow for the realization of photonic reservoirs consisting of hundreds or even thousands of small-scale lasers which can be coupled via external optical elements or integrated photonic structures. In addition to the small size footprint, microlasers offer the additional advantage that they operate in the regime of cavity quantum electrodynamics (cQED) which can reduce the threshold pump powers by orders of magnitude compared to conventional VCSELs to significantly improve the energy efficiency~\cite{Bjork1991,Reitzenstein2008}. Beyond that, quantum dot - micropillars can also act as bright electrically driven single-photon emitters~\cite{Heindel2010,Schlehahn2016} which provides exciting opportunities towards quantum rneural networks. In the following technological aspects and the fabrication of VCSELs, microlasers and single photon emitters for applications in photonic neuromorphic computing are discussed.    

\subsection{VCSEL fabrication}

The fabrication of VCSELs requires multiple nanoprocessing steps. It starts with the epitaxial growth of a planar microresonator structure by means of molecular beam epitaxy (MBE) or metal-organic chemical vapour deposition (MOCDV). As schematically shown in Fig.~\ref{fig:VCSEL_scheme} the microresonator is usually composed of a lower n-doped distributed Bragg reflector (DBR), the central cavity layer which is at least a $\lambda/2$ thick and includes the active area, and the upper p-doped DBR~\cite{Liu2015}. In the often used GaAs material system, the DBRs consists of typically more than 20 $\lambda/4$-thick AlGaAs/GaAs layers and the active medium in the central GaAs cavity is usually based on multiple InGaAs quantum wells located at the antinode of the confined light field~\cite{Michalzik2012}. In addition, the central cavity includes at least one thin AlGaAs layer with high Al-content (>90\%), which is later oxidized and acts as current window and leads to lateral light confinement governing the emission beam profile. The lateral nanoprocessing of VCSEL devices starts with the patterning of circular mesa structures with diameters in the range of 20 - 30 $\mu$m by means of UV lithography and plasma etching. Next, the current aperture with a well-defined inner diameter is formed by wet thermal oxidation of the cavity-integrated AlGaAs layer. Finally the lower n-contact and the upper ring-shaped p-contact are realized by additional UV lithography steps and metal deposition, leading to a VCSEL device as depicted in Fig.~\ref{fig:VCSEL_scheme}. It is noteworthy that especially in case of long wavelength VCSELs, intracavity contacts are preferred to avoid optical losses due to free-carrier absorption in the doped DBRs~\cite{Mehta2006}.. 

Beyond individual VCSELs, which have are today standard components for optical data communication, VCSEL arrays have become more and more important in recent years, finding applications for example in face recognition in modern cell phones~\cite{Liu2019}. Moreover, they are highly attractive to implement photonic reservoir computing and schemes relying on fast optical matrix multiplication using 1D VCSEL arrays~\cite{Hamerly2019}. While VCSEL arrays are in principle commercially available, neuromorphic applications have specific requirements which demand the development of customized VCSEL arrays with special geometry and optical properties. 

\begin{figure}
	\begin{center}
		\includegraphics[width=0.95\linewidth]{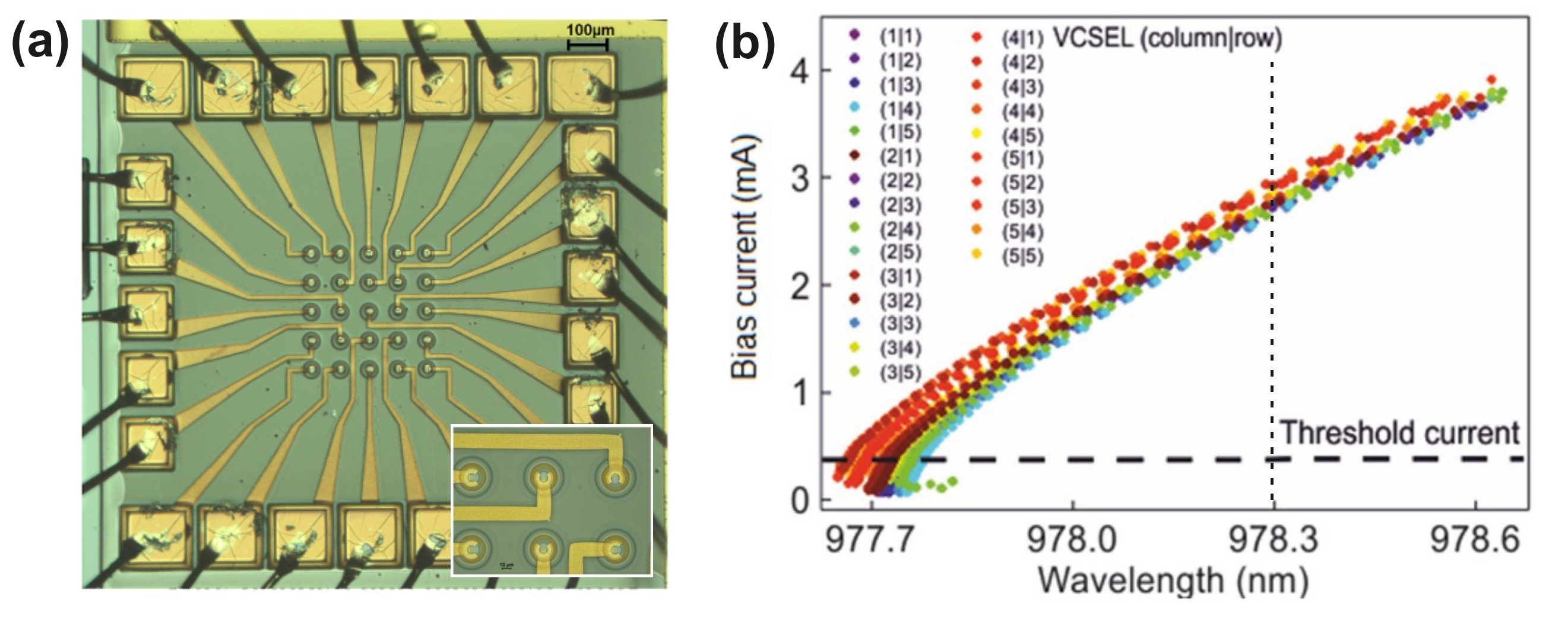}  
		\caption{(a) Optical microscope image of a compact 5 x 5 VCSEL array with a pitch of 80 $\mu$m. Each VCSEL is electrically contacted to allow for individual current injection and current-induced wavelength tuning. Inset: Zoom-in view of a subarray of 6 VCSELs. (b) Plot of the required injection currents to achieve a common target wavelength within the 5 x 5 VCSEL array. The dashed vertical line indicates that a common emission wavelength of 978.3 nm can be achieved by setting the injection currents of the individual emitters in a range of 2.6 - 3.0 $\mu$A. Reproduced with permission from Ref.~\cite{Heuser2020a}}
		\label{fig:VCSEL_array}
	\end{center}
\end{figure}

One example is a 5 x 5 VCSEL array optimized for photonic reservoir computing based on the diffractive coupling of VCSELs~\cite{Brunner2015}. Generally, neuromorphic computing concept require dense and spectrally homogenous VCSEL arrays, which in turn requires specific design and fabrication.
One example of such an array is shown in Fig.~\ref{fig:VCSEL_array}(a).
The 5$\times$5 array includes 25 VCSELs with a pitch of only 80~$\mu$m, which is significantly smaller compared to commercial arrays.
Dense packing facilitates efficient diffractive coupling within a available optical field of view~\cite{Maktoobi2020DiffCoupl}.
Additionally, the wafer material was chosen to guarantee a high spectral homogeneity, which is essential to allow for locking all VCSELs to an external injection laser.
As can be seen in Fig.~\ref{fig:VCSEL_array}(b), the emission wavelengths of the 25 VCSELs can be fine-tuned by the injection current of each individual  emitter~\cite{Heuser2020a}.
This enables a spectral homogeneity matching the injection locking range of 10-20 GHz. 
Moreover, due to a slight ellipticity of the emitters' cross-section, excellent polarization control of the 25 VCSELs is obtained with a standard deviation of only 1.5$^{\circ}$.

\subsection{High-$\beta$ microlaser and dense microlaser arrays}

Even VCSELs still pose limitations, not only in terms of the size-footprint, but also regarding the energy efficiency and modulation speed. To overcome such limitations, it is interesting to consider micro- and nanolasers which operate in the regime of cQED~\cite{Chow2018}. Here, small mode volumes and high cavity quality factors lead to enhanced light matter interaction which is quantified by the Purcell factor. The Purcell effect leads to a high fraction of spontaneous emission coupled into the lasing mode, which is described in terms of the $\beta$-factor and which leads to an order of magnitude reduction of the threshold pump power when comparing cavity-enhance high-$\beta$ micro- and nano-lasers with conventional semiconductor lasers~\cite{Bjork1991}. Additionally, the tight mode confinement in microlasers gives the opportunity to efficiently control the emission features such as the emission wavelength by the geometry of the laser cavity. 

Micropillar lasers with quantum dot (QD) gain medium are a popular type of microlasers~\cite{Gies2019}. Alike to VCSEL devices, they emit vertically, and they can be driven electrically in a straightforward way~\cite{Bckler2008}. In fact, large-scale micropillar arrays with small size footprint and high spectral homogeneity are ideally suited to realize the PRC proposed in Ref.~\cite{Brunner2015}. To meet the requirements of this neuromorphic computing scheme the QD-micropillar array has to feature a small pitch below 10 $\mu$m and high spectral homogeneity on a scale of the injecting locking range of a few tens of GHz, i.e. the spectral homogeneity of the array must be better than about 200 $\mu$V. 

\begin{figure}
	\begin{center}
		\includegraphics[width=0.85\linewidth]{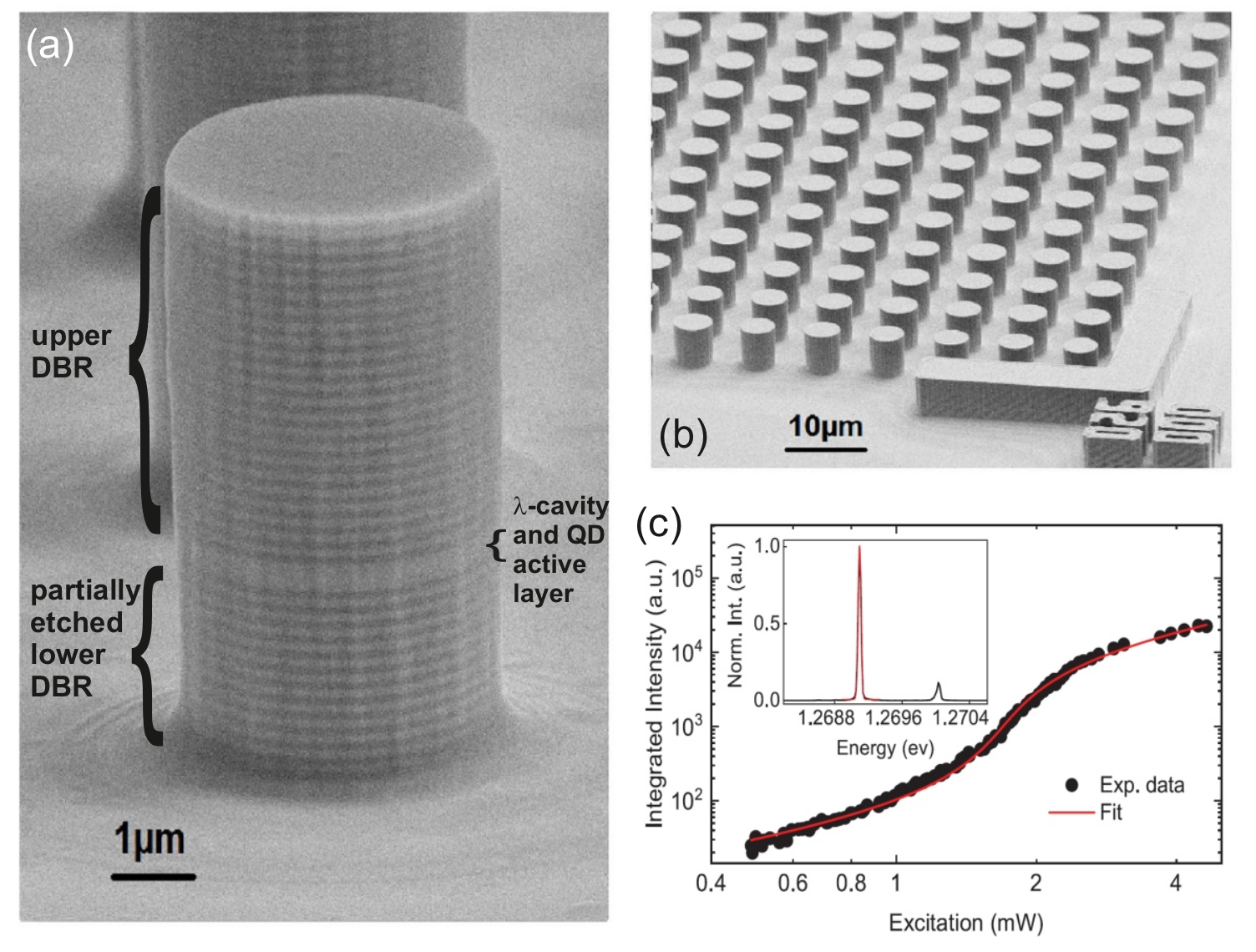} 
		\caption{Dense QD-micropillar arrays for photonic reservoir computing. (a) SEM image of a QD-micropillar with a diameter of 5 $\mu$m. (b) SEM image of a dense arrays of QD micropillars with a pitch of 8.3 $\mu$. (c) $\mu$PL emission spectrum (inset) and input-output characteristics of a 5 $\mu$m diameter micropillar with a $\beta$-factor of 2\%. Reproduced with permission from Ref.~\cite{Heuser2020IEEE}.}
		\label{fig:Pillar_SEM}
	\end{center}
\end{figure}

In order to fulfill the said requirements, the microlaser-arrays have to be fabricated by state-of-the-art nanotechnology platforms. Similar to VCSEL devices QD micropillar lasers are typically based on an epitaxial planar AlGaAs/GaAs microcavity structure with a lower and an upper DBR with up to 30 mirror pairs in an asymmetric design (3-4 more mirror pairs in the lower DBR to ensure directional emission via the upper DBR) and a central GaAs cavity, usually with a thickness of $\lambda$ and a single layer of self-assembled InGaAs QDs at the field antinode for optical gain. Fig.~\ref{fig:Pillar_SEM}(a) shows a scanning electron microscopy (SEM) image of a QD-micropillar with a lower (upper) DBR composed of 27 and (23) $\lambda$/4-thick Al$_{\mathrm{0.9}}$Ga$_{\mathrm{0.1}}$As/GaAs mirror pairs~\cite{Heuser2020IEEE}. The lower DBR is only partially (12 mirror pairs) etched, which is, however, sufficient to obtain the desired tight mode confinement, while maintaining a high Q-factor. Using electron beam lithography and plasma etching, dense arrays of such micropillar cavities can be realized with high accuracy. Fig.~\ref{fig:Pillar_SEM}(b) shows an excerpt of such an array consisting of 30 x 30 QD-micropillars with a pitch of 8.3 $\mu$m. 

An exemplary emission spectrum of an optically pumped QD-micropillar laser is depicted in the inset of Fig.~\ref{fig:Pillar_SEM}(c). It is dominated by emission of the fundamental pillar mode HE$_{\mathrm{11}}$ at about 1.269 eV. The input-output dependence of the fundamental emission mode is plotted in Fig.~\ref{fig:Pillar_SEM}(b) in double-logarithmic scale. The observed s-shaped behavior with a small non-linearity above the threshold pump power of about 1.8 mW is typical for cavity-enhanced microlasers. Fitting the experimental data with a rate-equations laser model allows us to extract a $\beta$-factor of 2\%. Higher $\beta$-factors close to unity can be obtained for QD-micropillars with smaller diameter and adiabatic microcavity design. 

Unfortunately, epitaxially grown micro-cavity wafers usually suffer from, both, a radial spectral dependence and local spectral inhomogeneities on a scale greater than what is acceptable for PRC. To address this issue and to reduce the spectral inhomogeneities of micropillar arrays advanced nanotechnology steps have to be applied for their fabrication. An interesting approach is ``diameter tuning`` which takes advantage of the thigh mode confinement in micropillar cavities to precisely adjust the emission energy of each pillar in the array via its diameter during fabrication to compensate wafer inhomogeneities~\cite{Heuser2018, Heuser2020IEEE}. There, the emission energy of the respective microcavity area is first recorded with $\mu$m accuracy by micro-photoluminesence ($\mu$PL) mapping relative to alignment markers (see lower right corner in Fig.~\ref{fig:Pillar_SEM}(b) for such a marker structure). With the knowledge of the local emission energie, the diameter of each micropillar is calculated where small (large) pillar diameters blue (red) shift its emission energy towards the target emission energy to compensate the spectral inhomogeneity of the underlying microcavity. As a result, after electron beam lithography and plasma etching, the spectral spread of emission energy of the micropillars in the fabricated arrays can be reduced from 1 meV to values as low as 120 $\mu$eV ~\cite{Heuser2020IEEE}. These results is very promising and further enhancement of the spectral homogeneity of large scale QD-micropillar arrays is expected by improved growth and nanoprocessing capabilities.

\section{Outlook \label{sec:Outlook}}

The application of standard and more advanced VCSELs in various neuromorphic computing concepts and schemes is by now an established approach, with clearly identified merits and future positibilities.
Similar to other areas in neuromorphic photonics, the crucuial future steps are now an increasingly seamless interfacing and integration with other techniques in order to form complete computing systems.
Crucial aspects here are (i) photonic integration, and (ii) more or less autonomous operation demonstrating a clear benefit in a relevant performance metric compared to other, and most importantly, to electronic neurmorphic computing concepts.

\section*{Funding}

The authors acknowledge the support of the Region Bourgogne Franche-Comt\'{e}. 
This work was supported by the EUR EIPHI program (Contract No. ANR-17-EURE- 0002), by the Volkwagen Foundation (NeuroQNet I\&II), by the French Investissements d’Avenir program, project ISITE-BFC (contract ANR-15-IDEX-03), partly by the french RENATECH network and its FEMTO-ST technological facility, and by the European Union’s Horizon 2020 research and innovation program under the Marie Skłodowska-Curie grant agreement No 713694 (MULTIPLY) and 860830 (POST DIGITAL).

\section*{Disclosures}
The authors declare no conflicts of interest.

\bibliography{biblio}
	
\end{document}